\documentclass[epj]{svjour}

\usepackage{epsfig,graphicx,bbm,psfrag,amssymb}
\usepackage{fancyhdr}
\def\makeheadbox{{
 }}
\def\mytitle{Lepton Flavour Violation in SUSY-seesaw: an update} 
\def\myauthors{Mar\'{\i}a J. Herrero}  
\def\mytype{Parallel Talk}
\def\mysession{Flavor Physics}

\psfrag{tanb}{\small $\tan{\beta}$}
\psfrag{tan b}{\small $\tan{\beta}$}
\psfrag{log(mN)}{\small $\log(m_N)$}
\psfrag{BR(tau -> 3mu)}{\small BR$(\tau \to 3 \mu)$}
\psfrag{BR(tau -> 3e)}{\small BR$(\tau \to 3 e)$}
\psfrag{BR(mu -> 3e)}{\small BR$(\mu \to 3 e)$}
\psfrag{BR(mu ->3e)}{\small BR$(\mu \to 3 e)$}
\psfrag{BR(tau -> mu + gamma)}{\small BR$(\tau \to \mu \gamma)$}
\psfrag{BR(tau -> e + gamma)}{\small BR$(\tau \to e \gamma)$}
\psfrag{BR(mu -> e + gamma)}{\small BR$(\mu \to e \gamma)$}
\psfrag{mod(theta2)}{\small $|\theta_2|$}
\psfrag{mod(theta1)}{\small $|\theta_1|$}

\begin{document}
\title{Lepton Flavour Violation in SUSY-seesaw: an update}
\author{
 Ernesto Arganda
\thanks{\emph{Email:} ernesto.arganda@uam.es}%
\and
 Mar\'\i a J. Herrero
\thanks{\emph{Email:} Talk given by maria.herrero@uam.es}%
}                     
%
%
\institute{Departamento de F\'{\i }sica Te\'{o}rica and Instituto de F\'{\i }sica 
Te\'{o}rica, IFT-UAM/CSIC,\\
Universidad Aut\'{o}noma de Madrid, Cantoblanco, E-28049 Madrid, Spain
}
%
\date{}
\abstract{
Here we update the predictions for lepton flavour violating
tau and muon decays, $l_j \to l_i \gamma$, $l_j \to 3 l_i$, and $\mu-e$ conversion in
nuclei. We work within a SUSY-seesaw context where the particle content of the 
Minimal Supersymmetric Standard Model is extended by 
three right handed neutrinos plus their corresponding 
SUSY partners, and where a seesaw mechanism for neutrino mass generation is implemented.
Two different scenarios with either 
universal or non-universal 
soft supersymmetry breaking Higgs masses at the
gauge coupling unification scale are considered. After comparing the predictions  
with present experimental bounds and future sensitivities, the most
promising processes are particularly emphasised. 
} 
\maketitle
%

\section{LFV within the SUSY Seesaw}
\label{intro}
The current knowlegde of neutrino mass differences and mixing angles clearly indicates
that lepton flavour number is not a conserved quantum number in Nature. However,
the lepton flavour violation (LFV) has so far been observed only in the neutrino sector.
One challenging task for the present and future experiments will then be to test if 
there is or there is not LFV in the charged lepton sector as well.


Here we focus in the Minimal Supersymmetric Standard Model (MSSM) enlarged by three right
handed neutrinos and their SUSY partners where potentially observable 
LFV effects in the charge lepton sector are expected to occur. 
We further assume a seesaw mechanism
for neutrino mass generation and use, in particular, 
the parameterisation proposed in~\cite{Casas:2001sr} where
the solution to the seesaw equation is written as 
$m_D =\,Y_\nu\,v_2 =\,\sqrt {m_N^{\rm diag}} R \sqrt {m_\nu^{\rm diag}}U^{\dagger}_{\rm
MNS}$. Here,
$R$ is defined by $\theta_i$
($i=1,2,3$);  $v_{1(2)}= \,v\,\cos (\sin) \beta$, $v=174$ GeV; 
$m_{\nu}^\mathrm{diag}=\, \mathrm{diag}\,(m_{\nu_1},m_{\nu_2},m_{\nu_3})$ denotes the
three light neutrino masses, and  
$m_N^\mathrm{diag}\,=\, \mathrm{diag}\,(m_{N_1},m_{N_2},m_{N_3})$ the three heavy
ones. $U_{\rm MNS}$ is given by
the three (light) neutrino mixing angles $\theta_{12},\theta_{23}$ and $\theta_{13}$, 
and three phases, $\delta, \phi_1$ and $\phi_2$. With this 
parameterisation is easy to accommodate
the neutrino data, while leaving room for extra neutrino mixings (from the right
handed sector). It further allows for large
Yukawa couplings $Y_\nu \sim \mathcal{O}(1)$ by
choosing large entries in $m^{\rm diag}_N$ and/or $\theta_i$. 

The particular LFV proccesses here studied are shown in table 1, together with their
present experimental bounds and future planned sensitivities. The predictions 
in the following are for two different constrained MSSM-seesaw scenarios, 
with universal and non-universal Higgs soft masses and
with respective parameters (in addition to the
previous neutrino sector parameters): 
1) CMSSM-seesaw: $M_0$, $M_{1/2}$, $A_0$ $\tan \beta$, and sign($\mu$), and 
2) NUHM-seesaw: $M_0$, $M_{1/2}$, $A_0$ $\tan \beta$, sign($\mu$),
$M_{H_1}=M_0(1+\delta_1)^{1/2}$ and
$M_{H_2}=M_0(1+\delta_2)^{1/2}$. All the predictions presented here include 
the full set of SUSY one-loop contributing diagrams, mediated by $\gamma$, Z,
 and Higgs bosons, as well as boxes, and do not use the Leading
 Logarithmic (LLog) nor the mass insertion approximations. 
This is a very short summary of the works in Refs.~\cite{Arganda:2005ji},~\cite{Antusch:2006vw} and~\cite{Arganda:2007jw} to which we
refer the reader for more details.  
\begin{table}[h]
\begin{center}
\caption{Present bounds and future sensitivities for the LFV 
processes.}
\begin{tabular}{|c|c  c |}
\hline
LFV process & Present bound & Future sensitivity \\
\hline
BR($\mu \to e\,\gamma$) & $1.2 \times 10^{-11}$  & $1.3 \times
10^{-13}$  \\
BR($\tau \to e \,\gamma$) & $1.1 \times 10^{-7}$ & 
 $10^{-8}$ \\
BR($\tau \to \mu \,\gamma$) & $4.5 \times 10^{-8}$  &
$10^{-8}$  \\
BR($\mu \to 3\,e$) & $1.0 \times 10^{-12}$  & 
$10^{-13}$  \\
BR($\tau \to 3\,e$) & $2.0 \times 10^{-7}$  & 
$10^{-8}$  \\
BR($\tau \to 3\,\mu$) & $1.9 \times 10^{-7}$  & 
$10^{-8}$  \\
CR($\mu - e$, Ti) & $4.3 \times 10^{-12}$  & 
$10^{-18}$  \\
CR($\mu - e$, Au) & $7 \times 10^{-13}$  & 
-  \\\hline
\end{tabular}
\label{LFV:bounds:future}
\end{center}
\end{table}
\section{Results and Discussion}
\label{results}

We focus on the dependence on the most relevant
parameters which, for the case of
hierarchical (degenerate) heavy neutrinos, are: the neutrino mass 
$m_{N_3}$ ($m_N$), $\tan\beta$, $\theta_1$ and $\theta_2$. We also study  
the sensitivity
of the BRs to $\theta_{13}$. The other input seesaw 
parameters $m_{N_1}$, $m_{N_2}$ and $\theta_3$, play a secondary role
since the BRs do not strongly depend on them. The light neutrino parameters are
fixed to: $m_{\nu_2}^2= \Delta m_{\rm sol}^2  +  m_{\nu_1}^2$,  
 $m_{\nu_3}^2= \Delta m_{\rm atm}^2  +  m_{\nu_1}^2$, 
$\Delta m^2_{\rm sol}=8\times 10^{-5}\,{\rm eV}^2$,
$\Delta m^2_{\rm atm}=2.5\times 10^{-3}\,{\rm eV}^2$,
$m_{{\nu}_1}=10^{-3}\,{\rm eV}$,
$\theta_{12}=30^\circ$, 
$\theta_{23}=45^\circ$,
$\theta_{13}\lesssim 10^\circ$ and
$\delta=\phi_1=\phi_2=0$.
\begin{figure}
\begin{center}
\psfig{file=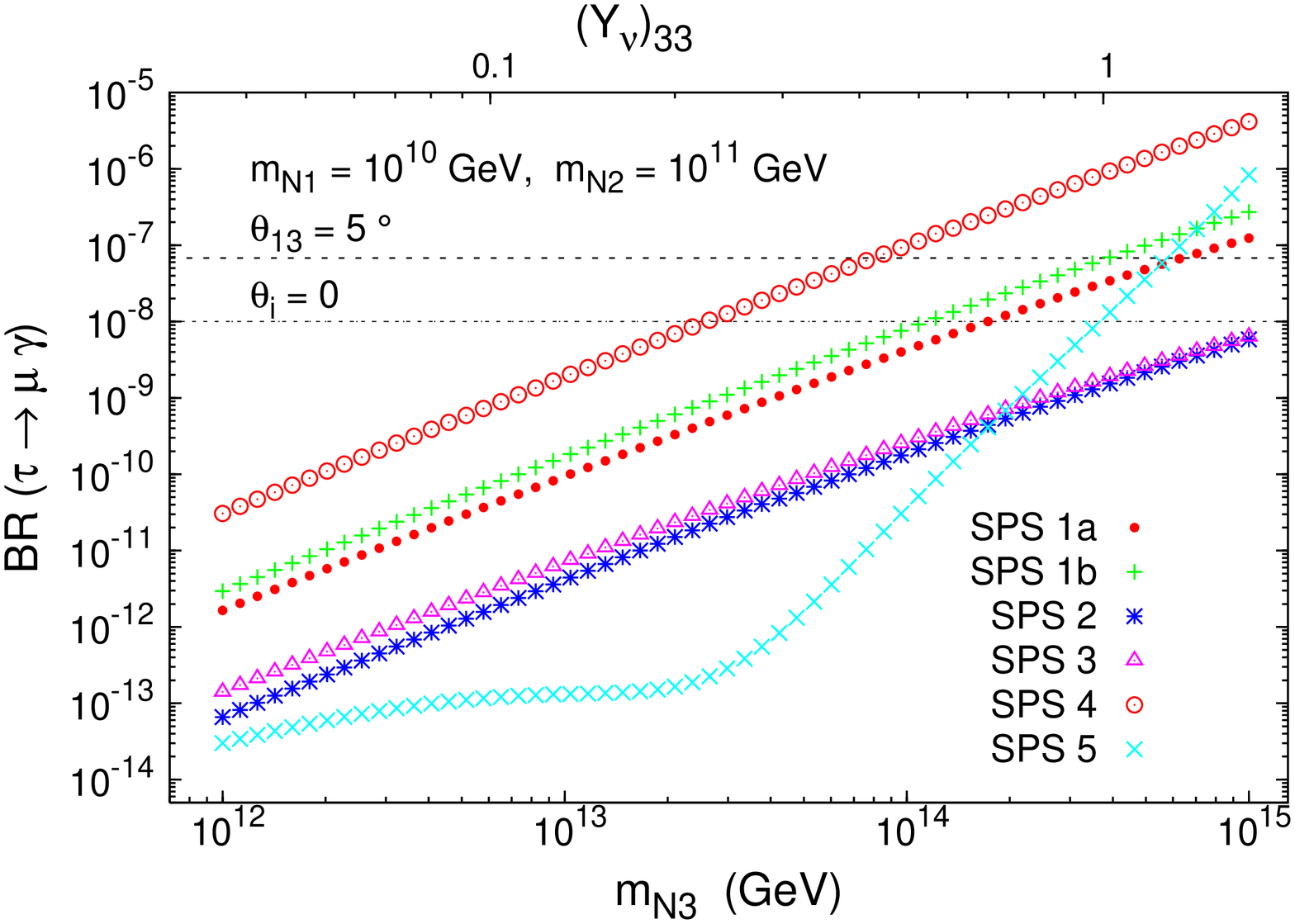,width=50mm,angle=270,clip=}
\psfig{file=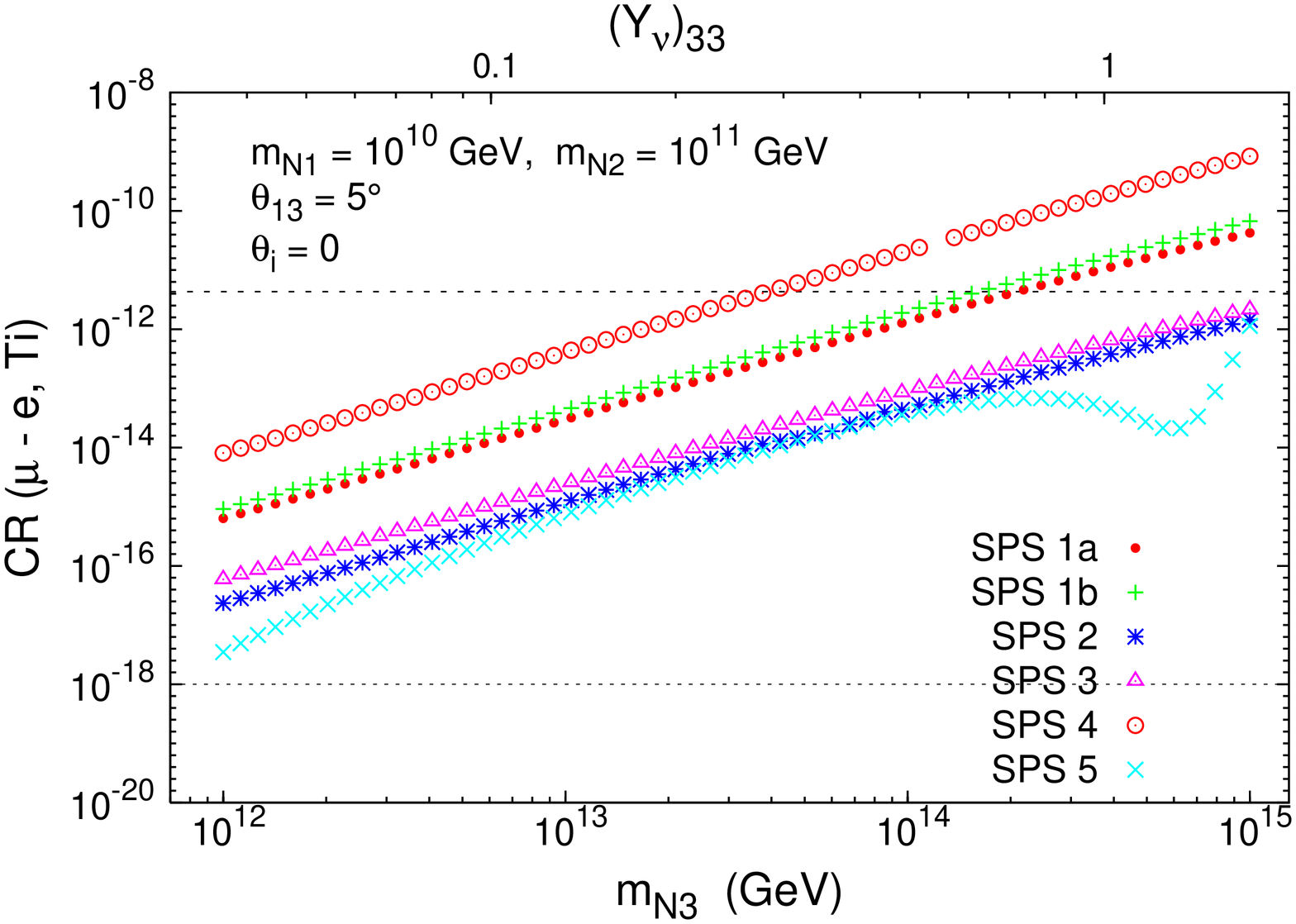,width=50mm,angle=270,clip=}
\caption{$\tau \to \mu \gamma$ and CR($\mu -e$, Ti) as a function of $m_{N_3}$.
      The predictions for SPS 1a
      (dots), 1b (crosses), 2 (asterisks), 3 (triangles), 4 (circles) and 5
      (times) are included. On the upper horizontal axis we display the associated 
      value of $(Y_\nu)_{33}$.
      In each case, we set $\theta_{13}=5^\circ$, and
      $\theta_i=0$. The upper (lower)
      horizontal line denotes the present experimental bound (future
      sensitivity).}
\label{fig:1}       
\end{center}
\end{figure}

\begin{figure}
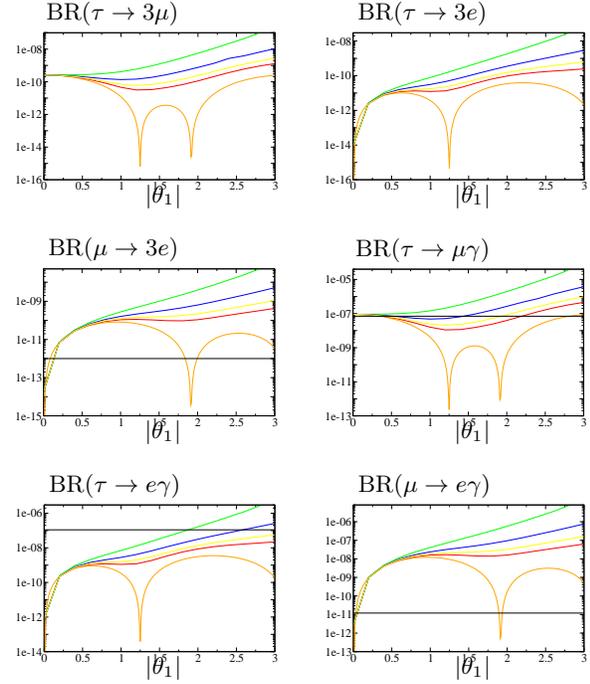

\begin{center}
\begin{tabular}{cc}
\psfig{file=figs/BRtau3mu_argtheta1.epsi,width=25mm,angle=270,clip=}
& \hspace*{1mm}
\psfig{file=figs/BRtau3e_argtheta1.epsi,width=25mm,angle=270,clip=}\\ \\
\psfig{file=figs/BRmu3e_argtheta1.epsi,width=25mm,angle=270,clip=}
& \hspace*{1mm}
\psfig{file=figs/BRtaumugamma_argtheta1.epsi,width=25mm,angle=270,clip=}\\ \\ 
\psfig{file=figs/BRtauegamma_argtheta1.epsi,width=25mm,angle=270,clip=}
& \hspace*{1mm}
\psfig{file=figs/BRmuegamma_argtheta1.epsi,width=25mm,angle=270,clip=}
\end{tabular}
\caption{Dependence of LFV $\tau$ and $\mu$ decays with 
$\vert \theta_1 \vert$ for SPS 4 case with $\arg(\theta_1) = 0, \pi/10, \pi/8, \pi/6, \pi/4$ in radians
(lower to upper lines), $(m_{N_1},m_{N_2},m_{N_3})=(10^8, 
2 \times 10^8, 10^{14})$ GeV, $\theta_2=\theta_3=0$, $\theta_{13}=0$ and
$m_{\nu_1}= 0$. The horizontal lines are 
the present experimental bounds.}
\label{fig:2} 
\end{center}
\end{figure}

\begin{figure}
\begin{center}
\psfig{file=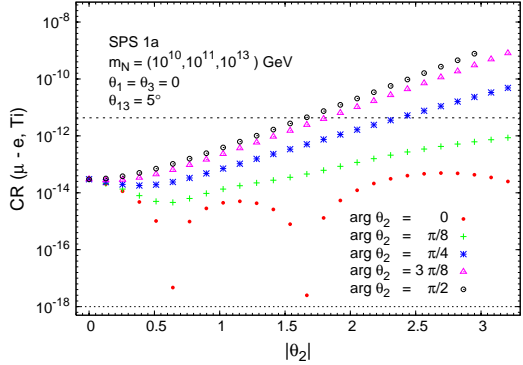,width=50mm,angle=270,clip=}
\caption{CR($\mu -e$, Ti) as a function of 
      $|\theta_2|$, for SPS 1a case
      with $\arg \theta_2\,=\,\{0,\, \pi/8\,,\,\pi/4\,,\,3\pi/8, \,\pi/2\}$ 
      (dots, crosses, asterisks, triangles and circles,
      respectively), $m_{N_i} =
      (10^{10},10^{11},10^{13})$ GeV, $\theta_{13}=5^\circ$.
      The upper (lower) horizontal line denotes the present
      experimental bound (future sensitivity).}
\label{fig:3}       
\end{center}
\end{figure}

\begin{figure}
\begin{center}
\psfig{file=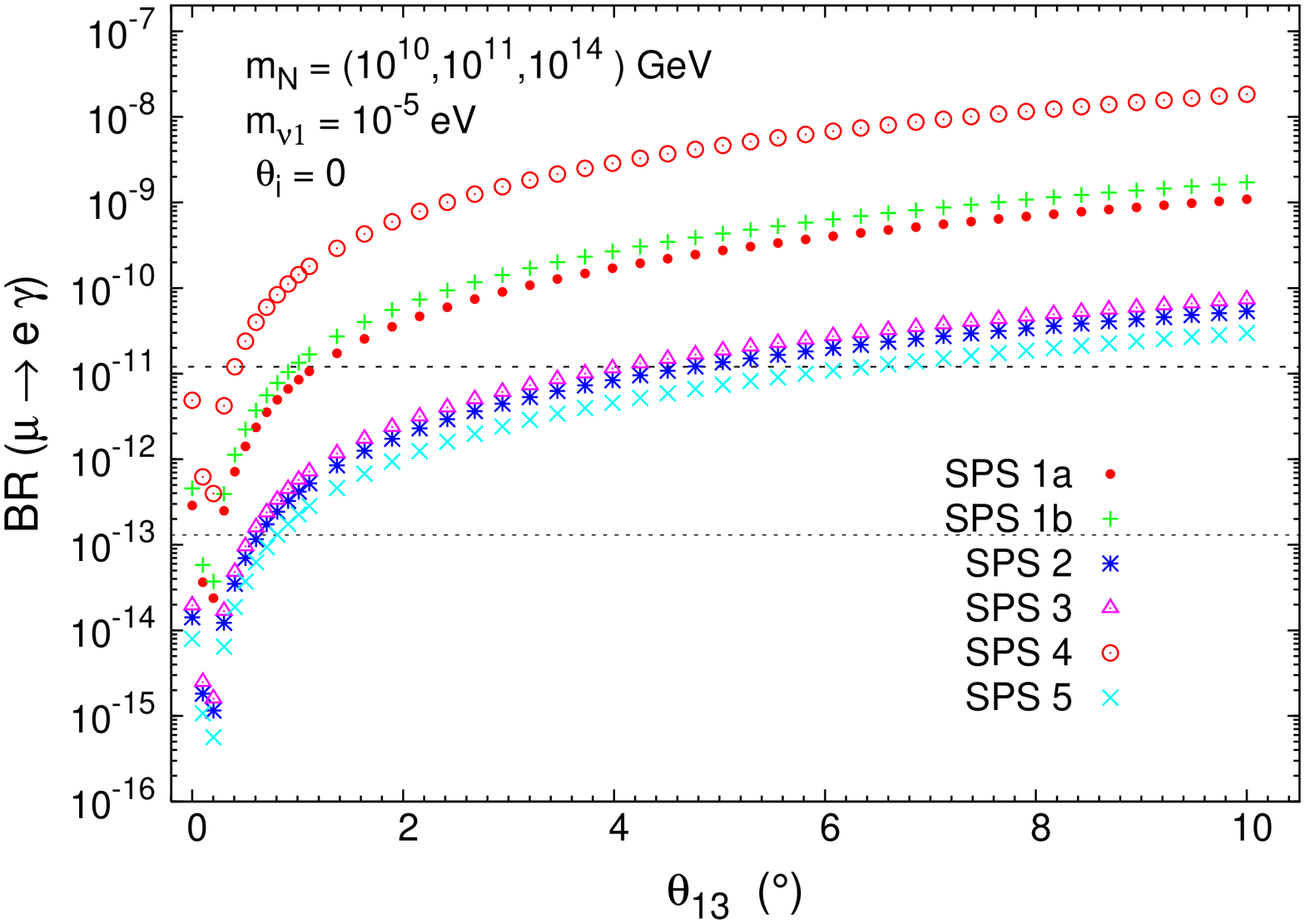,width=50mm,angle=270,clip=}
\psfig{file=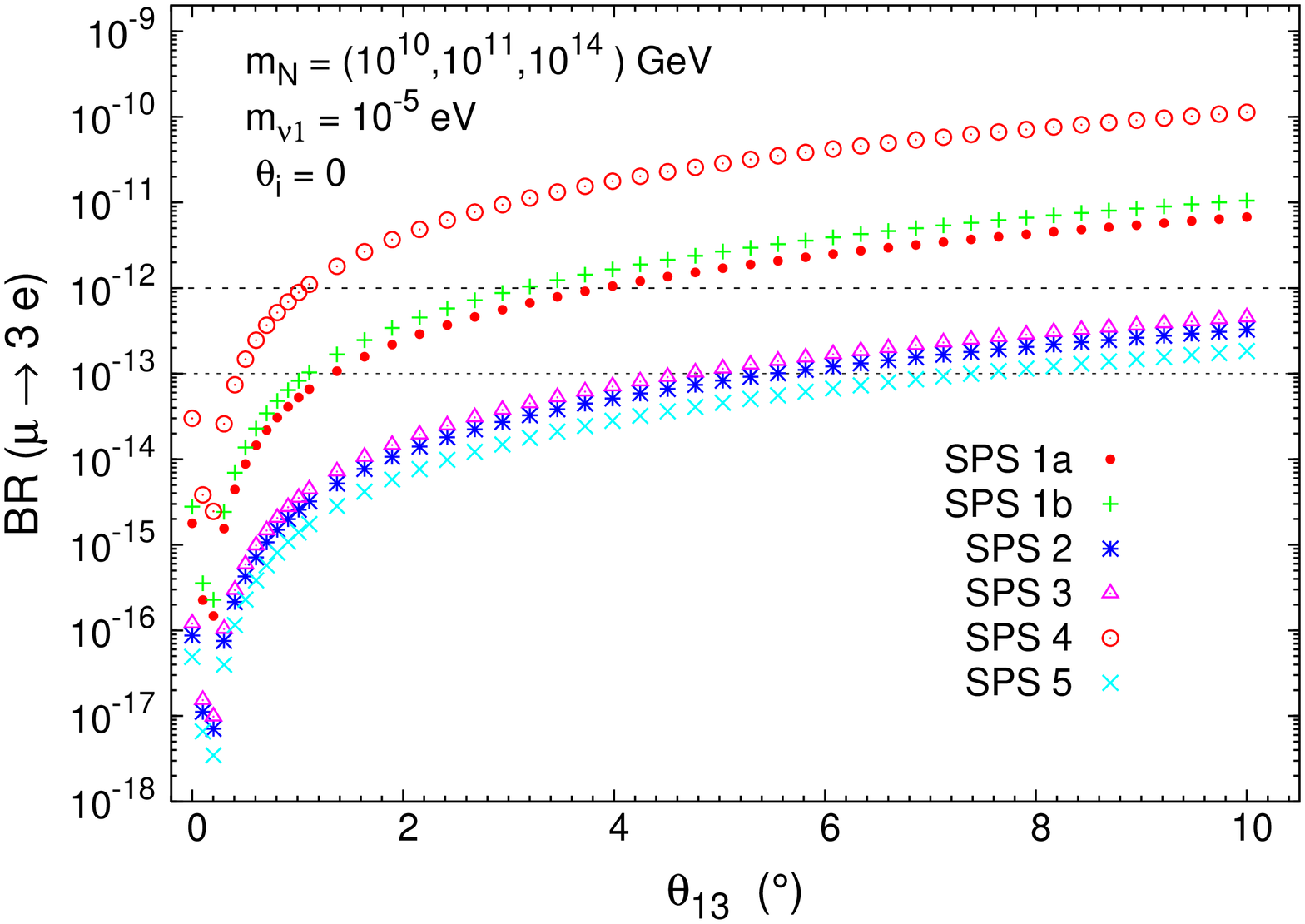,width=50mm,angle=270,clip=}
\psfig{file=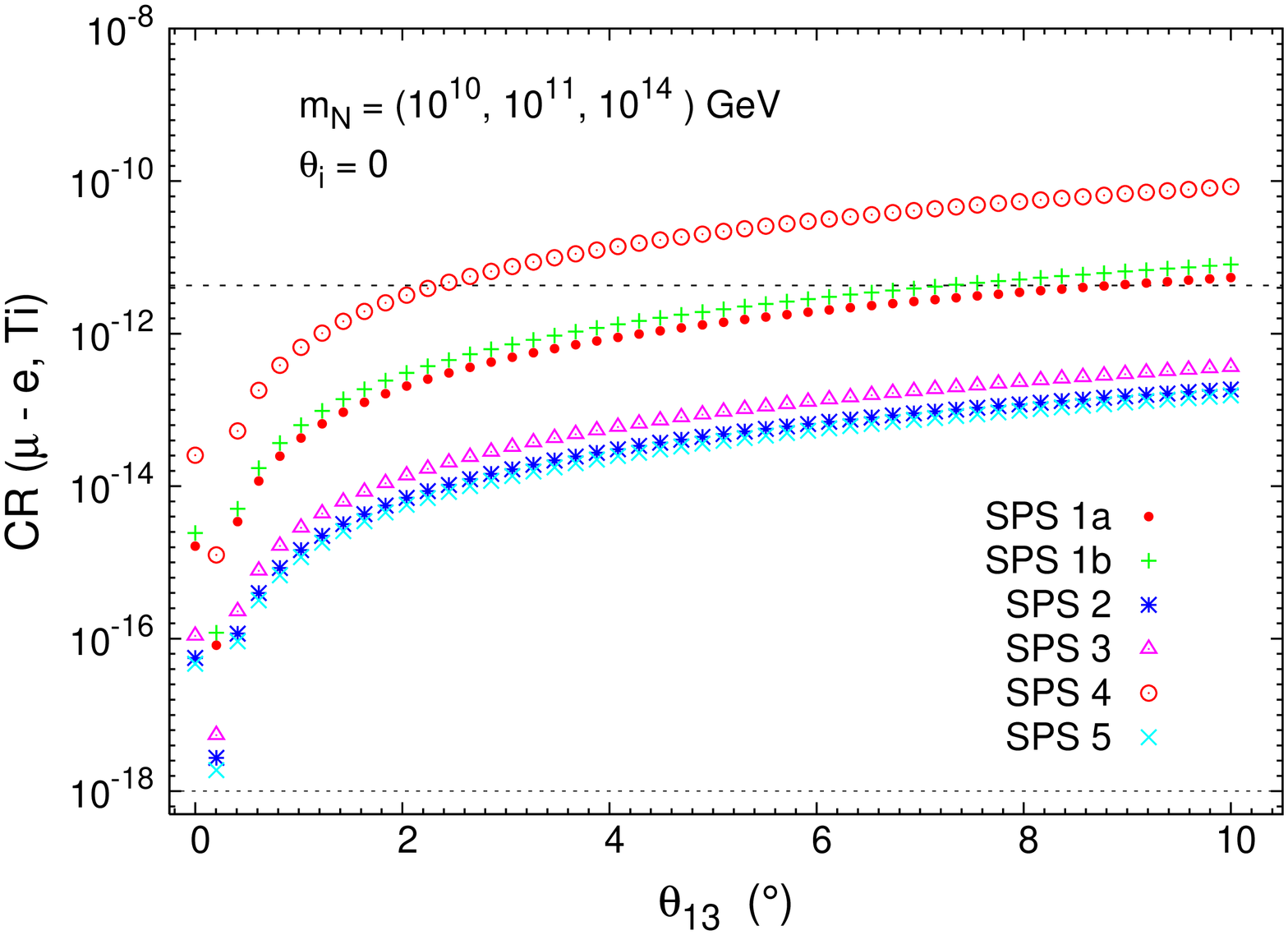,width=50mm,angle=270,clip=}
\caption{BR($\mu \to e \gamma$), BR($\mu \to 3e$) and CR($\mu - e$, Ti)
      as a function of $\theta_{13}$ (in
      degrees), for SPS 1a
      (dots), 1b (crosses), 2 (asterisks), 3 (triangles), 4 (circles) and 5
      (times), with
      $\theta_i=0$ and $m_{N_i} =
      (10^{10},10^{11},10^{14})$ GeV. The upper (lower)
      horizontal line denotes the present experimental bound (future
      sensitivity).}
\label{fig:4} 
\end{center}
\end{figure}

\begin{figure}
\begin{center}
\psfig{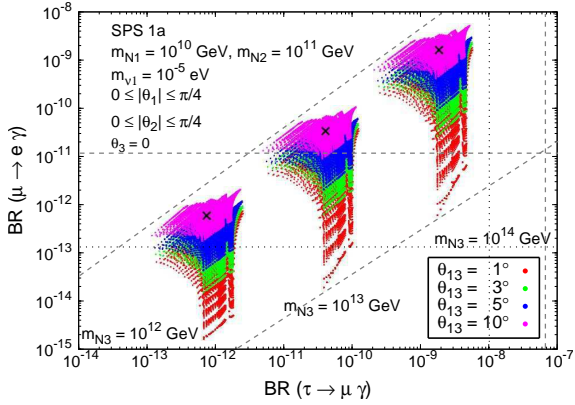}
\caption{Correlation between BR($\mu \to e\,\gamma$) and 
      BR($\tau \to \mu\,\gamma$) as a function of $m_{N_3}$, for SPS
      1a, and impact of $\theta_{13}$. The areas displayed represent the scan over $\theta_i$. From bottom to top, 
      the coloured regions correspond to 
      $\theta_{13}=1^\circ$, $3^\circ$, $5^\circ$ and $10^\circ$ (red,
      green, blue and pink, respectively). Horizontal and vertical 
      dashed (dotted) lines denote the experimental bounds (future
      sensitivities).}
\label{fig:5}
\end{center}
\end{figure}

The results for the CMSSM-seesaw scenario are collected in Figs.~\ref{fig:1}
through~\ref{fig:5}.
In Fig.~\ref{fig:1}, we display 
the predictions of BR$(\tau \to \mu \gamma)$ and CR($\mu -e$, Ti) as a function 
of the heaviest neutrino mass $m_{N_3}$ 
for the various SPS points~\cite{Allanach:2002nj},
and for the particular choice $\theta_i=0$ ($i=1,2,3$) 
and $\theta_{13}=5^\circ$. We have also considered the case of degenerate 
heavy neutrino spectra (not shown here).
In both scenarios for degenerate and hierarchical heavy neutrinos, we find 
a strong dependence on the the heavy neutrino masses, with the expected 
behaviour $~|m_N \log m_N|^2$ of the LLog approximation, except for SPS 5 point,
which fails by a factor of $\sim 10^4$. The rates for 
the various SPS points exhibit the following hierarchy,
BR$_{4}$~$>$~BR$_{\rm 1b}$~$\gtrsim$~BR$_{\rm 1a}$~$>
$~BR$_{3}$~$\gtrsim$~BR$_{2}$~$>$~BR$_{5}$. 
This behaviour can be understood in terms of the growth of the BRs with $\tan
\beta$, and from the different mass spectra associated with each point. Most of 
the studied processes reach their experimental limit at 
$m_{N_3} \in [10^{13}, 10^{15}]$ 
which corresponds to $Y_\nu^{33,32} \sim 0.1 - 1$.  At present, the most 
restrictive one is $\mu \to e \gamma$ (which sets bounds for SPS 1a of 
$m_{N_3} < 10^{13}-10^{14}$ GeV), although $\mu - e$ conversion will be the 
best one in future, with a sensitivity to $m_{N_3} > 10^{12}$ GeV.

Fig.~\ref{fig:2} shows the behaviour of the six considered LFV $\tau$ and $\mu$ 
decays, for SPS 4 point, as a function of $|\theta_1|$, for various values 
of arg$\theta_1$. We see clearly that the BRs for $0 < |\theta_1| < \pi$ and 
 $0 < {\rm arg} \theta_1 < \pi/2$ can increase up to a factor $10^2 - 10^4$ 
with respect to $\theta_i = 0$. Similar results have been found for $\theta_2$, 
while BRs are nearly constant with $\theta_3$ in the case of hierarchical 
neutrinos. The behaviour of CR($\mu -e$, Ti) with $\theta_i$ is very similar to 
that of BR($\mu \to e \gamma$) and BR($\mu \to 3e$). For instance, 
Fig.~\ref{fig:3} shows the dependence of CR($\mu -e$, Ti) with $\theta_2$,
and illustrates that for large $\theta_2$, rates up to a factor $\sim 10^4$ 
larger than in the $\theta_i=0$ case can be obtained. 

In Fig.~\ref{fig:4} we show
the dependence of $\mu \to e \gamma$, $\mu \to 3e$ and $\mu-e$
conversion on the light 
neutrino mixing angle $\theta_{13}$. These figures clearly manifest the 
very strong sensitivity
of their rates to the $\theta_{13}$ mixing angle for
hierarchical heavy neutrinos. Indeed, varying $\theta_{13}$ from 0 to 
$10^\circ$  
leads to an increase in the rates by as much as five orders
of magnitude. 
On the other hand, since $\mu \to e\,\gamma$ is very
sensitive to $\theta_{13}$, but 
BR($\tau \to \mu\,\gamma$) is clearly not, 
and since both BRs display the same approximate behaviour with 
$m_{N_3}$ and $\tan \beta$, one can study the impact that a potential
future measurement of $\theta_{13}$ and these two rates can have on the
knowledge of the otherwise unreacheable heavy neutrino parameters. The
correlation of these two observables as a function of $m_{N_3}$, is  
shown in Fig.~\ref{fig:5} for SPS~1a. 
Comparing
these predictions for the shaded areas along the expected diagonal
``corridor'', with the allowed experimental region, allows to conclude
about the impact of a $\theta_{13}$ measurement on the allowed/excluded 
$m_{N_3}$ values. The most important conclusion from Fig.~\ref{fig:5} is that for
SPS~1a, and for the parameter space defined in the caption, 
an hypothetical $\theta_{13}$ measurement larger than $1^\circ$, together 
with the present experimental bound on the BR($\mu \to e\,\gamma$),
will have the impact of excluding values of $m_{N_3} \gtrsim 10^{14}$
GeV. Moreover, with the planned MEG
sensitivity, the same $\theta_{13}$ measurement could further exclude 
$m_{N_3}  \gtrsim 3\times 10^{12}$~GeV.
\begin{figure}
\begin{center}
\psfig{file=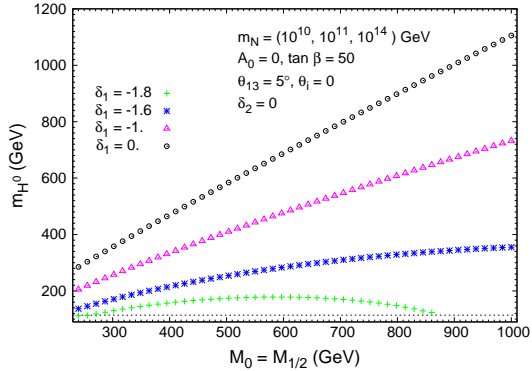,width=50mm,angle=270,clip=}
\caption{Mass of $m_{H^0}$
      as a function of $M_0=M_{1/2}$, 
      for fixed values of $\delta_1=\{-1.8,\,-1.6,\,-1,\,0\}$ 
      (respectively crosses, asterisks, triangles and circles),
      with $m_{N_i} = (10^{10},10^{11},10^{14})$
      GeV, $\theta_i=0$, $A_0=0$, $\tan \beta=50$ and
      $\theta_{13}=5^\circ$.}
\label{fig:6} 
\end{center}
\end{figure}

\begin{figure}
\begin{center}
\psfig{file=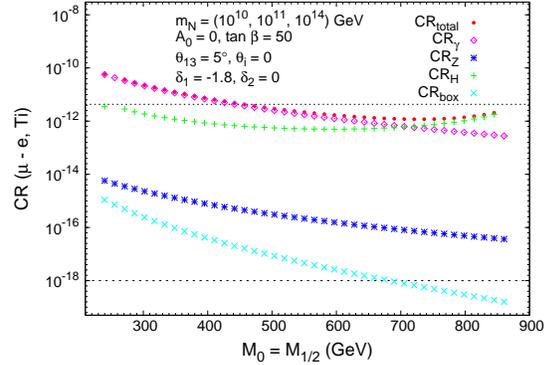,width=50mm,angle=270,clip=}
\caption{Contributions to CR($\mu -e$, Ti): 
    total (dots), $\gamma$-penguins (diamonds),
    $Z$-penguins (asterisks), $H$-penguins (crosses) and
    box diagrams (times) as a function of $M_0(=M_{1/2})$ for
    the NUHM case with $\delta_1 = -1.8$,
    $\delta_2 = 0$, $\tan \beta = 50$,
    $m_{N_i} = (10^{10},10^{11},10^{14})$
    GeV, $\theta_{13}=5^\circ$ and 
    $R=1$ ($\theta_i=0$). 
    The upper (lower) horizontal line denotes the present
    experimental bound (future sensitivity).}
\label{fig:7}
\end{center}
\end{figure}

\begin{figure}
\begin{center}
\psfig{file=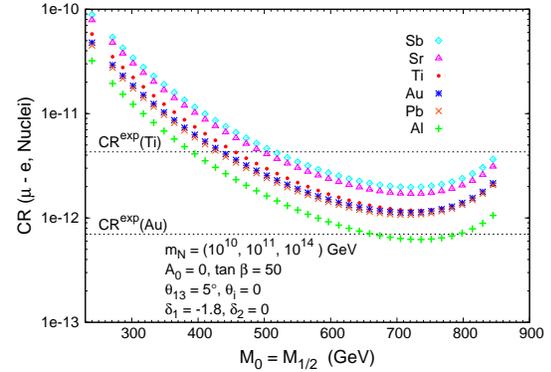,width=50mm,angle=270,clip=}
\caption{$\mu-e$ conversion rates as a function of
  $M_0=M_{1/2}$ in the NUHM-seesaw for various nuclei:
  Sb, Sr, Ti, Au, Pb and Al nuclei (diamonds,
  triangles, dots, asterisks, times and crosses, respectively),
  with $m_{N_i} =
  (10^{10},10^{11},10^{14})$ GeV, $A_0=0$, $\tan \beta =
  50$,  $\theta_{13}=5^\circ$, $\theta_i=0$, 
  $\delta_1=-1.8$ and $\delta_2=0$. From top to bottom, 
  the horizontal dashed lines denote the present experimental bounds
  for CR($\mu -e$, Ti) and CR($\mu -e$, Au).}
\label{fig:8}
\end{center}
\end{figure}

The numerical results for the NUHM-seesaw scenario as a function of 
$M_0=M_{1/2}=M_{\rm SUSY}$ are collected 
in Figs.~\ref{fig:6} through~\ref{fig:8}.
The behaviour of the predicted $m_{H^0}$ as a function of 
$M_{\rm SUSY}$ is shown in Fig.~\ref{fig:6}.
The most interesting solutions with important phenomenological
implications are found for negative $\delta_1$ and positive $\delta_2$. 
Notice that, for all the explored $\delta_{1,2}$ values,
we find a value of $m_{H^0}$ that is significantly smaller than in the 
universal case ($\delta_{1,2}=0$).

In Fig.~\ref{fig:7} 
the various contributions from the
$\gamma$-, $Z$-, Higgs mediated penguins and box diagrams as a function of 
$M_{\rm SUSY}$ are shown. Here, we choose $\delta_1 = -1.8$ and $\delta_2 = 0$.
We observe a very distinct behaviour with $M_{\rm SUSY}$ of the Higgs-mediated 
contributions compared to those of the CMSSM case. 
In fact, the Higgs-mediated contribution can
equal, or even exceed that of the photon,
dominating the total conversion rate in the large $M_0 = M_{1/2}$
region. 
These larger Higgs
contributions are the consequence of  
their exclusive SUSY non-decoupling behaviour 
for large $M_{\rm SUSY}$, and of 
the lighter Higgs boson mass values encountered in this region, 
as previously illustrated in Fig.~\ref{fig:6}.

In Fig.~\ref{fig:8} we
display the predicted $\mu-e$ conversion rates for other nuclei, 
concretely
Al, Ti, Sr, Sb, Au and Pb, as a
function of $M_{\rm SUSY}$. We clearly see that 
CR($\mu -e$, Sb) $>$  CR($\mu -e$, Sr)  $>$  CR($\mu -e$, Ti)  $>$
CR($\mu -e$, Au) $>$  CR($\mu -e$, Pb)  $>$  CR($\mu -e$, Al).
The most important conclusion from Fig.~\ref{fig:8} is
that we have found predictions for Gold nuclei which, for the
input parameters in this plot, are above its present experimental bound
throughout the explored $M_{\rm SUSY}$ interval.
Finally, althought not shown here for shortness,
we have also found an interesting loss of correlation between the 
predicted CR($\mu
-e$, Ti) and BR($\mu \to e \gamma$) in the NUHM-seesaw scenario compared to the 
universal case where these are known to be strongly correlated. 
This loss of correlation occurs when the
Higgs-contributions dominate the photon-contributions and 
could be tested if the announced 
future sensitivities in these quantities are reached. 

In conclusion, we believe that a joint measurement of the LFV branching ratios, the
$\mu - e$ conversion rates, 
$\theta_{13}$ and the sparticle spectrum will be a powerful tool for 
shedding some light on the otherwise unreachable heavy neutrino parameters.
Futhermore, in the case of a NUHM scenario, it may also provide 
interesting information on the Higgs sector.  
It is clear from this study that the connection between LFV and neutrino
physics will play a relevant role for the searches of new physics.

{\it We aknowledge} Ana M. Teixeira and Stefan Antusch for their participation 
in our previous works. M.J. Herrero thanks the organizors for her invitation
to this fruitful conference. Finantial support from the Spanish MEC, via 
AP2003-3776 and FPA2006-05423, and from the 'Comunidad de Madrid', via HEPHACOS
is also aknowledged.     

%

\end{document}